\documentclass[fleqn,usenatbib]{mnras}
\usepackage{subfigure}
\usepackage{footnote}
\usepackage[T1]{fontenc}
\usepackage{ae,aecompl}
\usepackage{graphicx}	
\usepackage{amsmath}	
\usepackage{amssymb}	
\usepackage{txfonts}
\usepackage{refcount}
\usepackage{float}

   \title{Properties of Sub-Neptune Atmospheres: TOI-270 System}

\author[J. Chouqar et al.]{
J. Chouqar,$^{1}$\thanks{E-mail: jamila.chouqar@ced.uca.ma}
Z. Benkhaldoun,$^{1}$
A. Jabiri,$^{1}$
J. Lustig-Yaeger,$^{2,3}$
A. Soubkiou$^{1}$
\newauthor
A. Szentgyorgyi$^{4}$
\\\\
$^{1}$Oukaimeden Observatory, PHEA Laboratory, Cadi Ayyad University, BP 2390 Marrakech Morocco\\
$^{2}$Department of Astronomy and Astrobiology Program, University of Washington, Box 351580, Seattle, Washington 98195, USA\\
$^{3}$NASA NExSS Virtual Planetary Laboratory, Box 351580, University of Washington, Seattle, Washington 98195, USA\\
$^{4}$Center for Astrophysics | Harvard \& Smithsonian, 60 Garden Street, Cambridge, MA 02138, USA\\
}

\date{Accepted XXX. Received YYY; in original form ZZZ}

\pubyear{2020}

\begin{document}
\label{firstpage}
\pagerange{\pageref{firstpage}--\pageref{lastpage}}
\maketitle

\begin{abstract}

We investigate the potential for the James Webb Space Telescope (JWST) to detect and characterize the atmospheres of the sub-Neptunian exoplanets in the TOI-270 system. Sub-Neptunes are considered more likely to be water worlds than gas dwarfs. We model their atmospheres using three atmospheric compositions - two examples of hydrogen-dominated atmospheres and a water-dominated atmosphere. We then simulate the infrared transmission spectra of these atmospheres for JWST instrument modes optimized for transit observation of exoplanet atmospheres: NIRISS, NIRSpec and MIRI. We then predict the observability of each exoplanet's atmosphere. TOI-270c \& d are excellent targets for detecting atmospheres with JWST transmission spectroscopy, requiring only 1 transit observation with NIRISS, NIRSpec and MIRI; higher signal-to-noise (SNR) can be obtained for a clear H-rich atmosphere. Fewer than 3 transits with NIRISS \& NIRSpec may be enough to reveal molecular features. Water-dominated atmospheres require more transits. Water spectral features in water-dominated atmospheres may be detectable with NIRISS in 2 or 3 transits. We find that the detection of spectral features in a cloudy, H-rich atmosphere does not require integrations as long as those required for the water-dominated atmosphere, which is consistent with the differences in atmospheric mean molecular weight. TOI-270c \& d could be prime targets for JWST transit observations of sub-Neptune atmospheres. These results provide useful predictions for observers who may propose to use JWST to detect and characterize the TOI-270 planet atmospheres.
\end{abstract}




\section{Introduction}

The potential to study the atmospheres of super Earths and sub-Neptunes is of considerable importance because they bridge a mass range that appears intrinsically common in the exoplanet population, but there are no examples of planets in this mass range in the Solar System. The study of planets with masses approaching that of the Earth brings us one step closer to eventually characterizing truly Earth-like exoplanets. Unlike Jupiter-mass exoplanets, super-Earths are expected to have diverse atmospheric compositions. They may obtain their atmospheres from different sources: capture of nebular gases, degassing during accretion, and degassing from subsequent tectonic activity, that should strongly depend on the formation history and subsequent evolution of each individual planet \citep{sea, hu}. Determining the composition of super-Earth atmospheres has therefore become a priority for exoplanet observers and theorists alike. However, the initial observations of super-Earth atmospheres have revealed significant challenges to achieving this goal. The first challenge is the small size of super-Earths that makes them difficult to characterize with transit spectroscopy. A second issue that has risen to the forefront: clouds \citep[e.g.][]{Kreidberg2014}. Clouds make the atmosphere opaque at a given wavelength which masks the expected molecular and atomic absorption \citep{for3}.\\
The recently-launched Transiting Exoplanet Survey Satellite \citep[TESS; ][]{tess} is revolutionizing the field of exoplanet science by discovering a multitude of temperate Earths, super-Earths and sub-Neptunes in transiting orbits around bright stars in the solar neighborhood. Stars in the TESS catalogue are extremely bright, so that detailed characterizations of the planets and their atmospheres can be performed.\\ 
The TOI-270 planetary system is a prime example of the systems TESS was designed to discover. Since it is bright and at least three planets transit the host star, it is an interesting target from the perspective of atmospheric characterization. TOI-270 is a nearby, quiet M3V-type star, 22.5 parsec away, that hosts at least three planets - one super-Earth-sized (TOI-270b) and the two sub-Neptunes (TOI-270c \& d) (\citet{max}, see Tab. \ref{para} for details). Because TOI-270 is bright and quiet, we will likely be able to characterize the atmospheres of planets b, c \& d. Furthermore, these planets have low equilibrium temperatures and thus are very close to being potentially habitable (see Tab. \ref{para}). The information we get from the characterization will make it possible to discriminate among different formation models for small planets near the habitable zone. Systems like this are currently very rare. The mass diversity of the exoplanets in this system provides an interesting case study for planet formation and photo-evaporation. Thus, it makes it a prime target for further investigations with transmission spectroscopy. Therefore, the outer planets (TOI-270c \& d) may be two of the best planets suitable for atmospheric characterization with the upcoming JWST and ground-based ELTs.\\
Only three rocky planets in the Solar System have significant atmospheres. It seems inevitable that there are many exoplanet examples, which may be surrounded by atmospheres with different structures and compositions. Some of these planets may have high enough surface gravity \citep[e.g.][]{val, val1, sot, for2, sea2} to be able to retain large hydrogen-rich atmospheres, while others will bear a closer resemblance to Earth, with atmospheres depleted in hydrogen and composed of predominantly heavier molecules \citep[e.g.][]{elkin, eliza, kempton, marcy, roger, fluton}. \citet{zeng2} used a growth model and conduct Monte Carlo simulations to demonstrate that  many sub-Neptunes planets around sun-like stars are expected to be "water worlds". Depending on their formation and evolutionary history, super-Earths may have atmospheres rich in $H$/$He$, $H_{2}O$, and $CO_{2}$.\\ 
In this letter we explore the detectability of atmospheres and molecular species with JWST, given several different atmospheric composition and structure models. We model the observational signature of molecular constituents in these models for both sub-Neptunes TOI-270c and d using Pandexo \citep{bat}. We use the approach of \citet{jacob}, applied for the TRAPPIST-1 system \citep{Gillon2016, Gillon2017, Luger2017}, 
to see whether molecules might be detected by JWST in the atmospheres of either TOI-270c and TOI-270d. We describe our model atmospheres and we present the modeled transmission spectra for TOI-270c and d in Section 2. In Section 4 \& 5, we show results from JWST simulations and assess the observability of each planet. Finally, we summarize our results in Section 6. 
\section{Atmospheric Model}
\begin{table}
	 \small
	\centering   
	\caption{Global parameters of the TOI-270 System}
	\label{para}
	\begin{tabular}{cccc}  
		\hline \hline \\    
		Star :  Mag  & R ($R_{\odot}$) & $ M $ ($M_{\odot}$) & $ T_{eff}^{b}$($K$) \\ \\   
		V=12.62 \& J=9.099  & 0.38  & 0.40 & 3386 \\ 
		\hline \hline \\     
		Planets   & \textcolor{blue}{TOI-270b} & \textcolor{blue}{TOI-270c} &  \textcolor{blue}{TOI-270d} \\
		\hline \\             
		R ($R_{\oplus}$) & 1.247  & 2.42 & 2.13 \\
		\hline \\
		Period (d)  & - & 5.66 & 11.38  \\ 
		\hline \\
		Transit duration T(h) & - & 1.65 & 2.14  \\
		\hline
		\hline \\
		$ M^{a} $ ($M_{\oplus} - est$) & 1.9 & 6.6 & 5.4 \\
		\hline \\
        g ($ ms^{-2} $ -- est) & - & 11 & 11.7 \\		
		\hline \\ 
		$ T_{eq}^{b}$ ($K$) & -  & 463 & 372 \\
		\hline \\ 
		$ K$ ($ms^{-1}$) & - & 4.19 & 2.76 \\
		\hline \\                                       
	\end{tabular}
    \footnotesize{ \\ \textbf{Notes}. The stellar parameters \& planetary radius, periods and Transit duration were taken from \citet{max}.\\$^a$ Masses are estimated based on \citet{kip}\\ $^b$ The equilibrium Temperatures are calculated assuming a zero\\ albedo and energy redistribution over 4$ \pi $ sr.} 
    \end{table}
\begin{figure}
	\centering
	\includegraphics[width=\columnwidth]{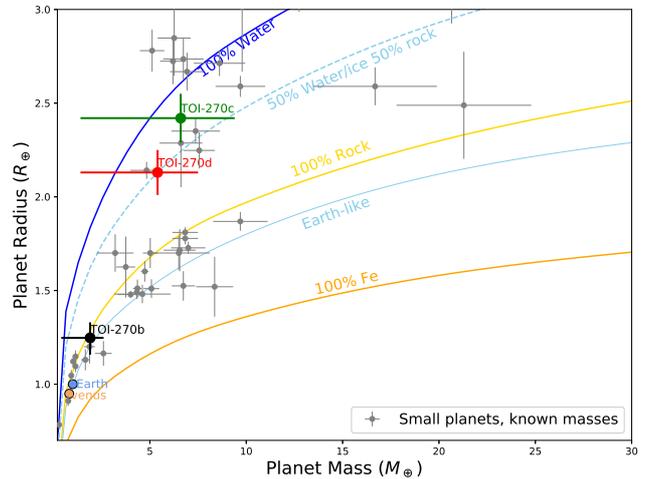}
	\caption{Mass-radius relationships for known exoplanets with mass-radius measurements better than 20\%; TOI-270b, c and d are identified in black, green and red resp. Composition models are from \citet{for2}. "Earth-like" here means a composition of 33\% Fe and 67\% $MgSiO_{3}$, whereas "100\% Rock" means a composition of 100\% $MgSiO_{3}$. Earth and Venus are identified in this plot as a pale blue and orange circles resp.}
	\label{mr}		
\end{figure}
\subsection{Planet Properties}
We take the majority of properties used to model the sub-Neptunes
planets from the discovery paper \citep{max}, including
planetary radii, stellar radii and temperatures, and orbital distance (see Tab. \ref{para}). The masses and radii of the planets are shown in Fig. \ref{mr} and compared with theoretical and empirical mass-radius relationship from \citet{for2}. The masses of these small planets have not yet been measured. Applying the probabilistic mass-radius relation of \citet{kip}, we obtained M $\approx$ $1.9^{+1.5}_{-0.7}$ $M_{\odot}$, M $\approx$ $6.6^{+5.2}_{-2.8}$ $M_{\odot}$ \& M $\approx$ $5.4^{+4.0}_{-2.1}$ $M_{\odot}$ for TOI-270b, c \& d respectively. These masses are consistent with the water-ice/gas-dominated compositions for TOI-270c \& d, whereas we see that the super-Earth TOI-270b likely falls into the regime of Earth-like/rocky compositions (Fig. \ref{mr}). \citet{kip} proposed a probabilistic mass-radius relation
conditioned on a sample of 316 well constrained objects to fit a continuous broken power-law model across a much larger range of masses and radii. Their code, Forecaster, is available to the community\footnote{\url{https://github.com/chenjj2/forecaster}} and is able to reproduce a larger spread in radius (or mass) than the previous power law relations.\\ 
Planets around bright M-dwarfs are amenable to precision radial velocity observations. The expected velocity amplitude of the stellar reflex motion caused by TOI-270c planet is 4.19 $ms^{-1}$ while TOI-270d planet will cause a RV motion of a 2.76 $ms^{-1}$ (Tab. \ref{para}). Several existing instrument can measure at the $\sim$ 1 $ms^{-1}$ level, especially the two HARPS-S \citep{Mayor} and ESPRESSO \citep{pepe}, which have the spectral resolution to observe the RV signals from each TOI-270 planet. Also the star is observable with JWST because it is so red and JWST is an IR mission. 


\subsection{Model Description}
\label{m.d}
We used the open source petitRADTRANS\footnote{\url{http://gitlab.com/mauricemolli/petitRADTRANS}} radiative transfer and retrieval code \citep{mol} to calculate synthetic transmission spectra for TOI-270c and d assuming two atmospheric scenarios : hydrogen-rich and water-dominated. petitRADTRANS, a radiative transfer Python package, consists of two resolution modes. The correlated-k mode (c-k) calculates spectra making use of the k approximation, at a wavelength spacing of $ \lambda$/$\Delta\lambda $ = 1000. This method is based on the fact that within a spectral band $\Delta\nu$, which is sufficiently narrow to assume a constant Planck function, the precise knowledge of each line position is not required for the computation of uniform column average transmissivities. The line-by-line mode (lbl) calculates the radiative transfer in a line-by-line fashion, that is directly in wavelength space. The line-by-line wavelength spacing is $\lambda$/$\Delta\lambda$ =$ 10^{6}$. Moreover, the code makes it possible to include clouds in different ways, by:
\begin{itemize}
	\item i) defining a cloud deck pressure, below which the atmosphere cannot be probed,
	\item ii) defining parametric descriptions for the wavelength dependence of the cloud opacity,
	\item iii) using cross-sections derived from optical constants of various condensates. The cloud opacities of the species are calculated assuming either homogeneous and spherical, or irregularly shaped cloud particles.
\end{itemize}
 
In this work, one model scenario we explore is that of sub-Neptunes with a primordial H/He-rich atmosphere accreted from the protosolar nebula. This nebula is assumed to have a composition roughly similar to that of the Sun. The second scenario is a water-world planet with a sublimating H2O envelope (water-rich).
\textit{Hydrogen-Rich Scenarios}. We consider that both planets are surrounded by a clear isothermal hydrogen-rich atmosphere (mean molecular weight $\mu$ = 2.39). We use an isothermal atmosphere because we do not have sufficient information to justify a more complex structure. Furthermore transmission spectra, unlike thermal emission spectra, are not highly sensitive to vertical temperature gradients in a planet's atmosphere. They are, however, sensitive to the absolute temperature of the atmosphere \citep{for3, kemp1}. Modeling transmission spectra with isothermal P-T profiles has been shown to be sufficient to explain observations \citep[see][]{for3, heng, atmo, kemp1}. However, we note that the higher quality data expected from JWST may justify using transmission spectrum retrievals that explicitly fit for the vertical TP structure, particularly for hot Jupiters \citep{roch}, although the significance of the TP structure for interpreting the spectra of smaller super-Earths, like those in the TOI-270 system (see also, \cite{kemp1}), remains a key open question \citep{barstow2020}. For each planet, we calculate the equilibrium temperature assuming zero albedo and full heat redistribution. These temperatures are reported in Tab. \ref{para}. We model the spectra by including $H_{2}O, CH_{4}, CO, CO_{2}$ and $NH_{3} $, as well as $ H_{2} - He$ and $H_{2} - H_{2}$ collision-induced absorption, as opacity sources in the atmosphere. Rayleigh scattering is included for $H_{2}$. We refer the reader to \citet{mol} for the most up-to-date sources for the line lists. We select the k mode to calculate the combined opacity of the mixture (at low resolution $ \lambda$/$\Delta\lambda $ = 1000). The chemical equilibrium mixing ratios are calculated with the Chemical Equilibrium with Applications (CEA) model \citep{Gor, lin, lin1, viss, mos, lin2}. CEA minimizes the Gibbs free energy with an elemental mass balance constraint given a local temperature, pressure, and elemental abundances. We include species that contain H, He, C, O and N.

\textit{Cloudy H-Rich Scenarios}. The petitRadtrans code allows placement of an opaque cloud layer at a pressure level of our choice. In this work, we examine the effects of clouds by
placing an opaque cloud deck at a pressure of 0.01 bar for
both planets with hydrogen-dominated atmospheres.

\textit{Water-Dominated Scenarios}. We assume that the planets are enveloped in a clear, isothermal water-dominated atmosphere composed of 95\% $H_{2}O$ and 5\% $CO_{2}$. The model includes the $H_{2}O$ and $CO_{2}$ Rayleigh scattering cross-sections. We also compare it to a pure water-planet (100\% $H_{2}O$) with $H_{2}O$ Rayleigh scattering.

%

\subsection{Transmission Spectra}

\begin{figure}
	\begin{minipage}{0.98\linewidth} 
		\centering
		\includegraphics[width=\columnwidth]{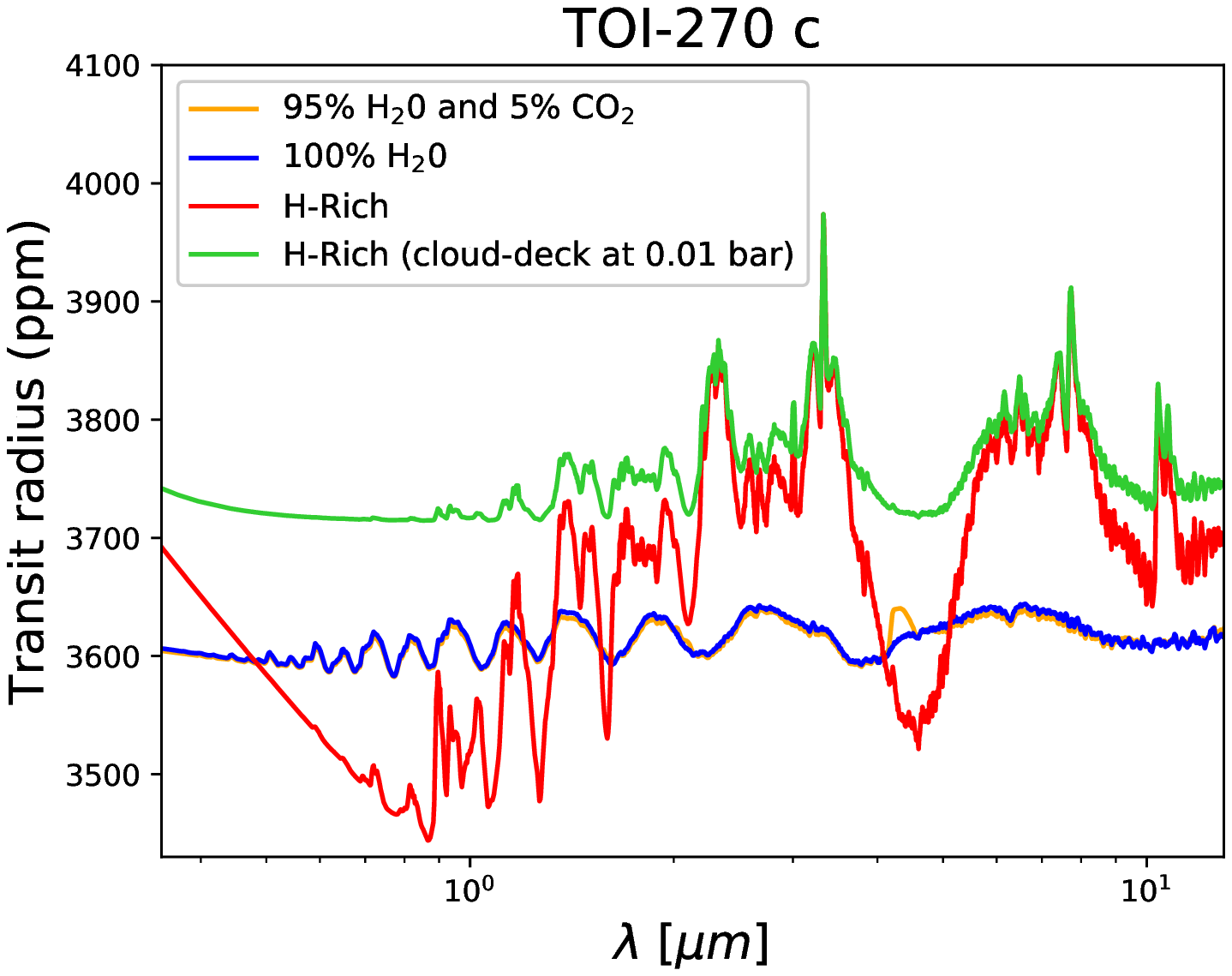}
	\end{minipage}
	\begin{minipage}{0.98\linewidth} 
		\vspace{0.2cm}
		\centering
		\includegraphics[width=\columnwidth]{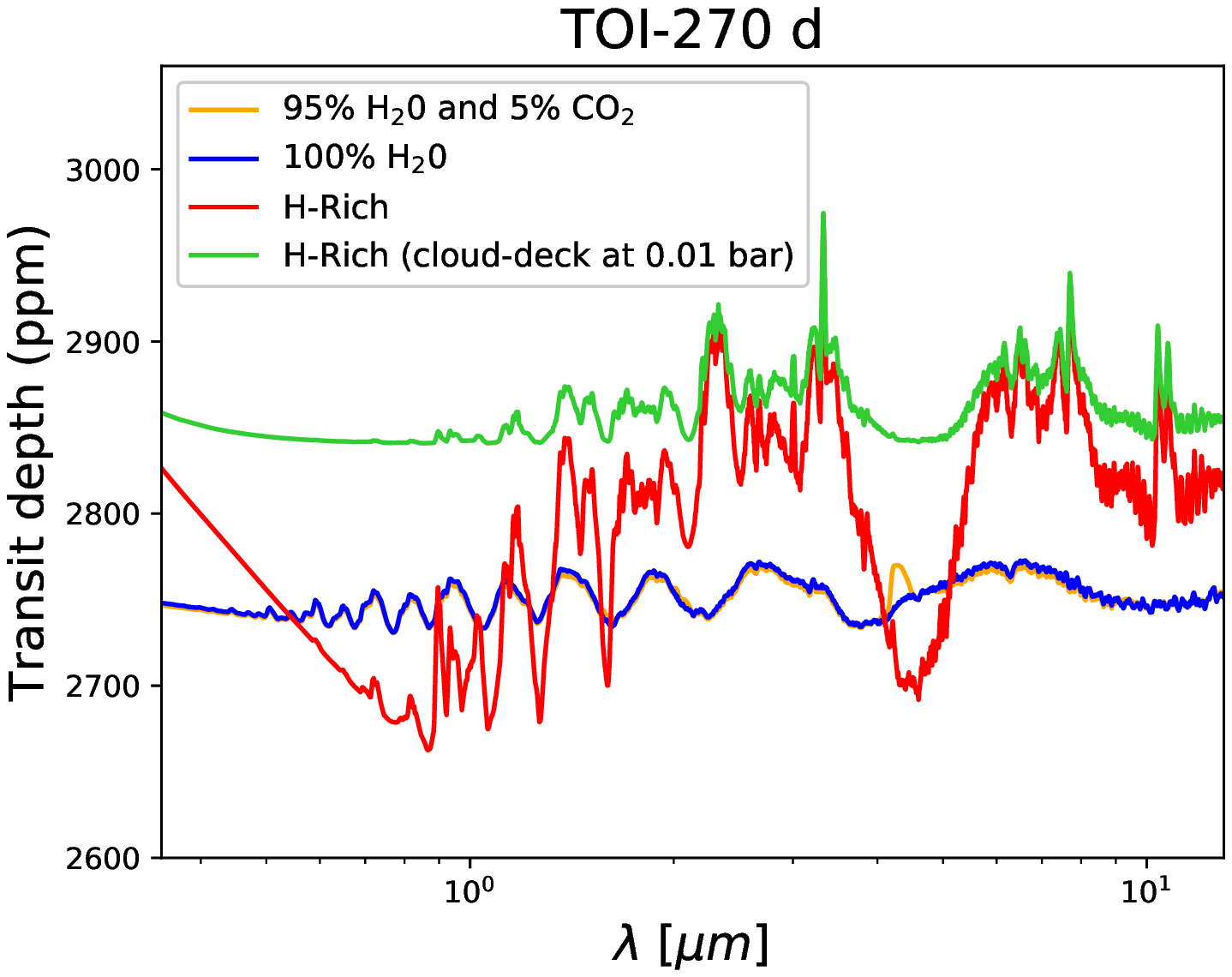}
	\end{minipage}
	\caption{Forward model transmission spectra of TOI-270c (Top panel) and TOI-270d (Bottom panel). The red line is a hydrogen rich atmosphere with a solar metallicity, the green curve is for an atmosphere that is hydrogen rich with a solar metallicity and a cloud layer at 0.01 bar, the orange line is for an atmosphere that is rich of water (95\% $H_{2}O$ and 5\% $CO_{2}$) and the blue curve is for an atmosphere that is dominated by water (100\% $H_{2}O$). The models shown are smoothed for clarity.}
	\label{spectre}		
\end{figure}
The modeled transmission spectra resulting from these calculations are shown in Fig. \ref{spectre} for each of the four cases of atmospheric composition that were discussed in Section \textcolor{blue}{\ref*{m.d}}. In all model planets, we set the 1 bar radius equal to the transit radius observed. The transmission spectrum of TOI-270c and d will be dominated by the combined feature of $H_{2}O$, $CH_{4}$, and $NH_{3}$ in the near-infrared wavelengths if the planets have an $H_{2}$-dominated cloud-free atmosphere. An atmosphere with 95\% $H_{2}O$ and 5\% $CO_{2}$ ($H_{2}O$/$CO_{2}$ = 19) can be distinguished from an atmosphere with 100\% $H_{2}O$  because the transit depth within the $CO_{2}$ band at 4.5 $\mu$m would be higher relative to the transit depths in the $H_{2}O$ bands, as shown in Fig. \ref{spectre}.\\
The strength of the atmospheric features is dependent on the atmospheric scale height H, which in turn is inversely proportional to the atmosphere's mean molecular weight. For scenarios discussed above, the mean molecular weight could plausibly vary from $\mu$ = 2 for an atmosphere dominated by molecular hydrogen to $\mu$ = 18 for atmospheres dominated by heavier molecules like $H_{2}O$. As can be seen in Fig. \ref{spectre}, the hydrogen-rich atmosphere shows deep absorption features in transmission at a level of 356 and 332 ppm for TOI-270c and TOI-270d respectively. The cloudy H-rich atmosphere has weaker features, 0.6 and 0.4 times weaker than that found for clear atmosphere for both TOI-270c and d respectively. This is because the presence of clouds in planetary atmospheres blocks transmission of stellar radiation above some pressure level. This restricts our ability to detect spectral features due to molecular absorption other than in regions of the atmosphere above the level of the clouds deck. The spectral features are even weaker for the water-dominated model due to its small scale height. This means that the transmission spectrum can only probe a very narrow range of heights in the atmosphere. 


\subsection{C/O effects in H-Rich and Non H-Rich Atmosphere} 

\begin{figure}
	\begin{minipage}{0.98\linewidth} 
		\centering
		\includegraphics[width=\columnwidth]{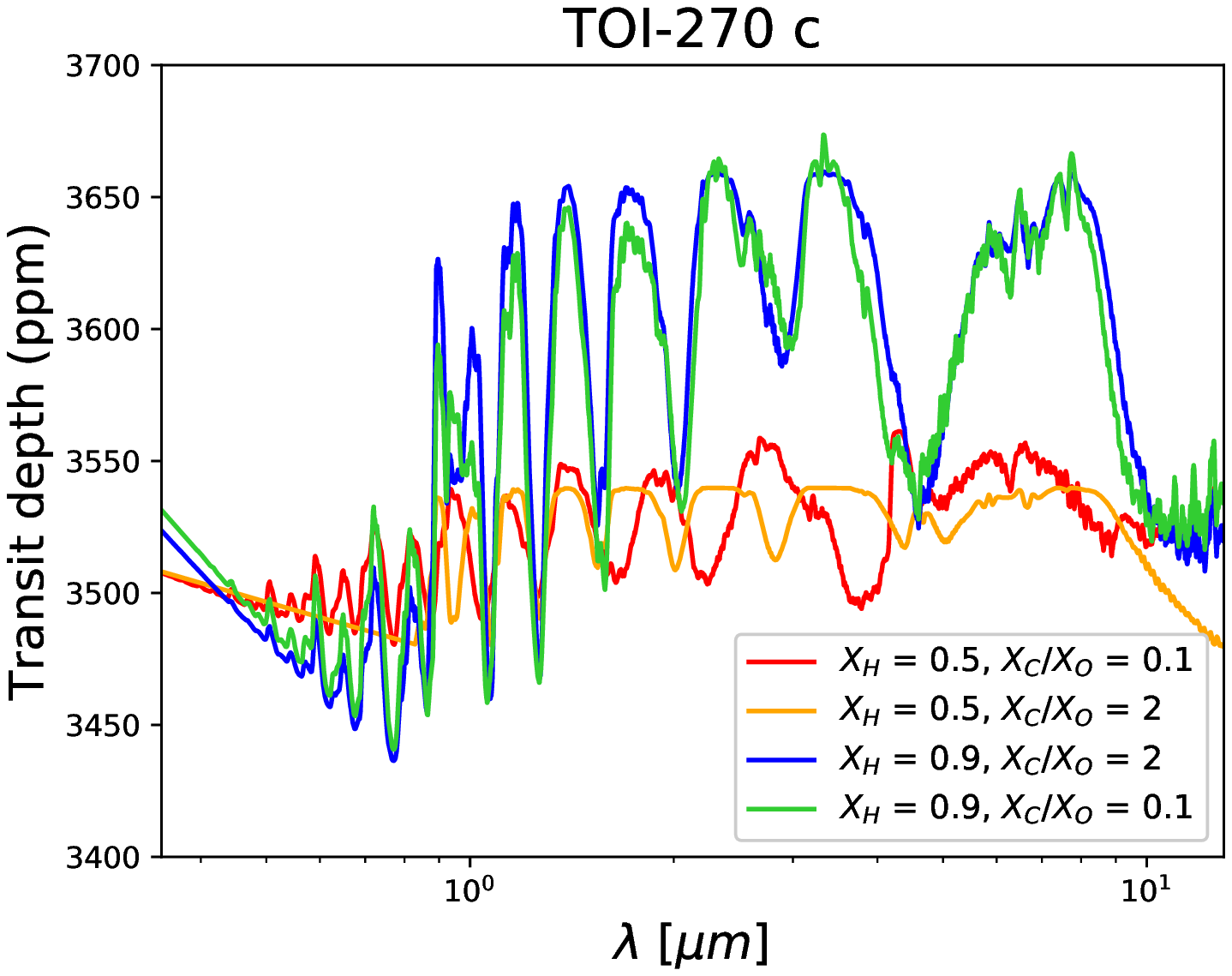}
	\end{minipage}
	\begin{minipage}{0.98\linewidth} 
		\vspace{0.2cm}
		\centering
		\includegraphics[width=\columnwidth]{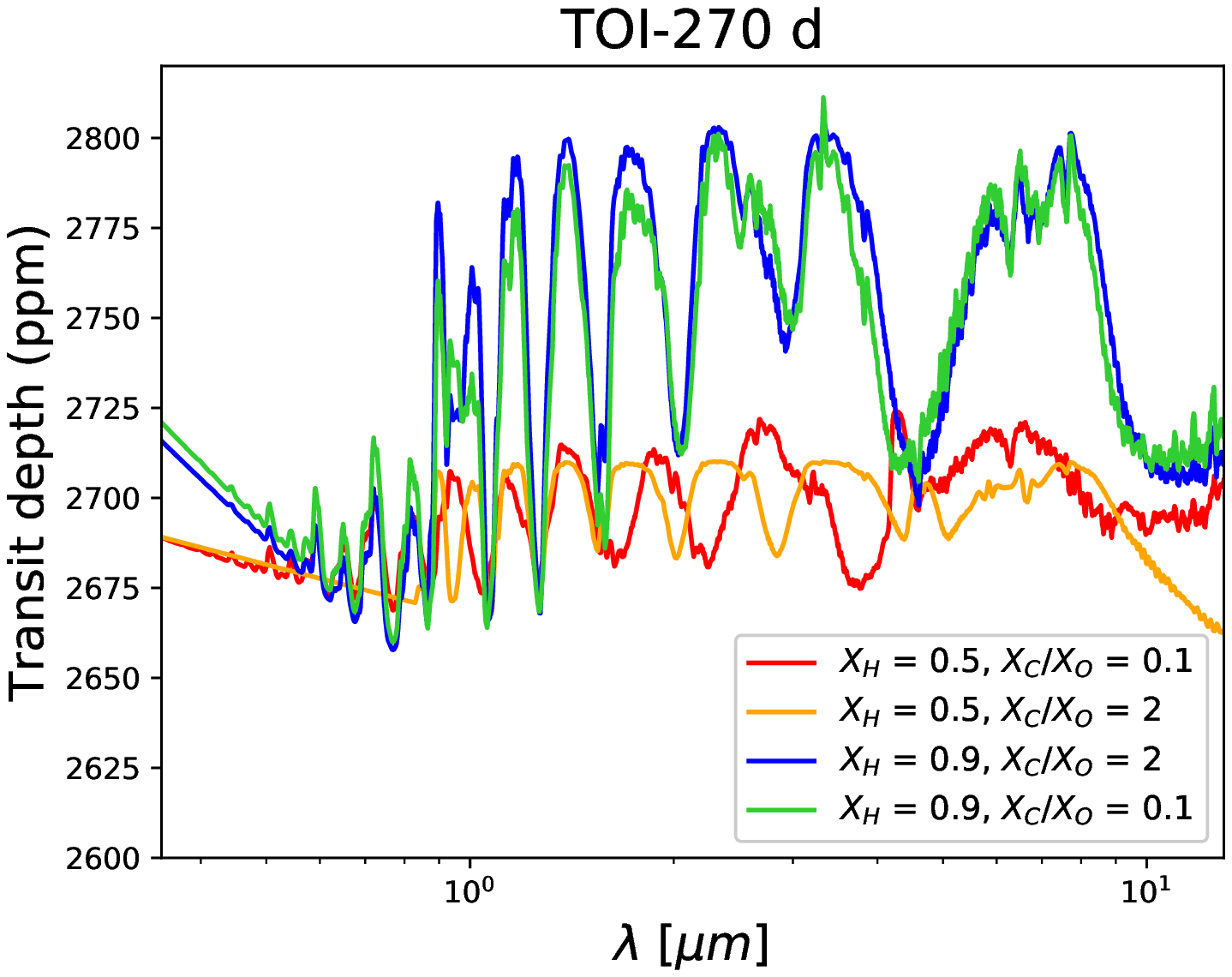}
	\end{minipage}
	\caption{ Modeled transmission spectra for TOI-270c and TOI-270d. The presented spectra are for both $H_{2}$-dominated ($X_{H}$ = 0.9) and non-$H_{2}$-dominated ($X_{H}$ = 0.5) atmospheres with a variety of carbon to oxygen ratios ranging from oxygen-rich to carbon-rich. For each $X_{H}$, we show the spectra corresponding to the carbon to oxygen ratio of 0.1 and 2.}
	\label{co}
\end{figure}

In this Section, we simulate thick atmospheres for exoplanets with a variety of C-H-O elemental abundances. We focus on the C-H-O chemistry as they are the most common elements in the universe. We explore different C/O ratios ranging from oxygen-rich ($X_{C}$/$X_{O}$ = 0.1) to carbon-rich ($X_{C}$/$X_{O}$ = 2) for each clear H-rich ($X_{H}$ = 0.9) and non-H-rich ($X_{H}$ = 0.5) atmosphere. Transmission spectra of TOI-270c and d are displayed in Fig. \ref{co} to further guide future observations. For a H-rich atmosphere the chemical equilibrium calculations result in a composition of mostly $H_{2}$, along with significant quantities of $H_{2}O$ and $CH_{4}$. Carbon can be present as $CO_{2}$,$CO$ or $CH_{4}$ in a planet's atmosphere, as well as other, minor species. In an oxidized atmosphere irradiated by intense stellar UV, $CH_{4}$ has a short half-life and is rapidly converted to $CO$ or $CO_{2}$ by UV photolysis followed by a reaction with other atmospheric species. By contrast, in an atmosphere with $> 70\%$ hydrogen, methane is extremely long lived. This is because photolysis of methane produces $CH_{3}$ radicals, which in an $H_{2}$-dominated atmosphere, predominantly react with hydrogen atoms or molecules to regenerate 
 $CH_{4}$ \citep{hu, sea3}. If the atmosphere is non-H-rich, the dominant molecules produced by C-H-O can be $H_{2}O$, $CH_{4}$, $CO_{2}$ and $CO$.\\ 
As shown in Fig. \ref{co}, it will be difficult to distinguish a carbon-rich atmosphere versus an oxygen-rich atmosphere since the methane is the dominant form of carbon, and water is the dominant form of oxygen in the $H_{2}$-dominated atmosphere, thus the mixing ratios of $CH_{4}$ and $H_{2}O$ only have linear dependency on $X_{C}$/$X_{O}$. However, a carbon-rich atmosphere can be uniquely identified by detecting the absorption bands of $CH_{4}$ at 0.9 $\mu$m and 1 $\mu$m (Fig. \ref{co}). If the planets have a non-H-rich atmosphere, we see that the size of the features in the transmission spectrum are quite small (Fig. \ref{co}). Therefore, it is not possible to discriminate between a carbon-rich atmosphere and an oxygen-rich atmosphere on the basis of transmission spectra for TOI-270c or d. It may be possible to distinguish them using thermal emission spectroscopy \citep[see][]{hu}.
\section{Atmospheric Characterization}
\begin{figure}
	\centering
	\includegraphics[width=\columnwidth]{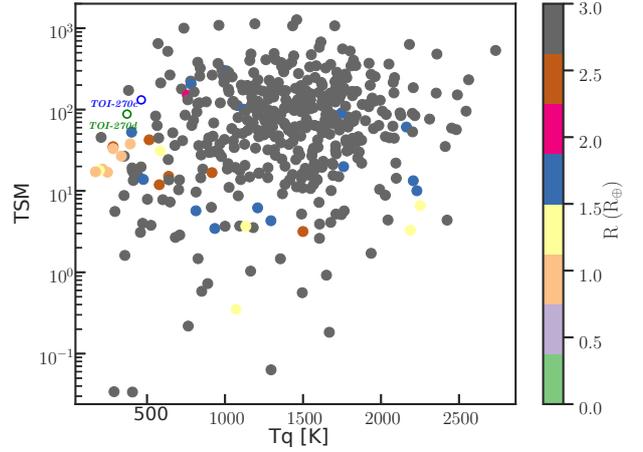}
	\caption{The transmission spectroscopy metric as defined by \citet{kem} for all exoplanet candidates with $ R_{p} $ < 3 $R_{\oplus}$ as a function of planetary equilibrium temperature. The sample of small exoplanets is taken from the NASA Exoplanet Archive$ ^{\textcolor{blue}{1}} $. The color represents the radius of the planet, The TOI-270 planets are the empty colored symbols.}
	\label{tsm}
\end{figure}

The brightness and the quietness of the M3V TOI-270 host star immediately argues that any transiting planet will be attractive for atmospheric characterization. Observations of a planetary atmosphere through transmission spectroscopy during transit provide opportunities to measure the extent, kinematics, abundances, and structure of the atmosphere \citep{sea1}. Such measurements can be utilized to address fundamental questions such as planetary atmospheric escape and interactions with the host star
\citep{cau, Arney2017}, formation and structure of planetary interiors \citep{owen}, planetary and atmospheric evolution \citep{ober, Meadows2018}, and biological processes \citep{mead, Schwieterman2018}.\\
\citet{kem} outlined a way to prioritize transiting planets for transmission spectroscopy via a "transmission spectroscopy metric" (TSM) roughly equal to the SNR for JWST's NIRISS instrument. Using the mass and radius determined in this work for each planet, we obtained a transmission metric value of  131 and 90 for the sub-Neptunes TOI-270c and d respectively, which have a radius within the range 1.5 $R_{\oplus}$ <$ R_{p} $ < 2.75 $R_{\oplus}$ (see Tab. \ref{para}). A higher TSM means a higher S/N, so it would be easy to characterize the TOI-270c \& d atmospheres with the NIRISS instrument. Figure \ref{tsm} shows the TSM as a function of equilibrium temperature for known small exoplanets with $ R_{p} $ < 3 $R_{\oplus}$. The sample of small exoplanets is taken from the NASA Exoplanet Archive \footnote{\url{https://exoplanetarchive.ipac.caltech.edu}}. 
Out of this sample of super-Earth exoplanets and with these values of the TSM, that represent the expected signal-to-noise ratio of transmission spectroscopy observations with JWST, both planets in the TOI-270 system are well suited targets for atmospheric characterization. This is mainly a consequence of the brightness of this nearby cool, small, star. 
\section{Predictions For JWST Observations}
\begin{figure*}
	\begin{minipage}{1\linewidth}
		\begin{center}
			\subfigure{%
				
				\includegraphics[width=0.5\textwidth]{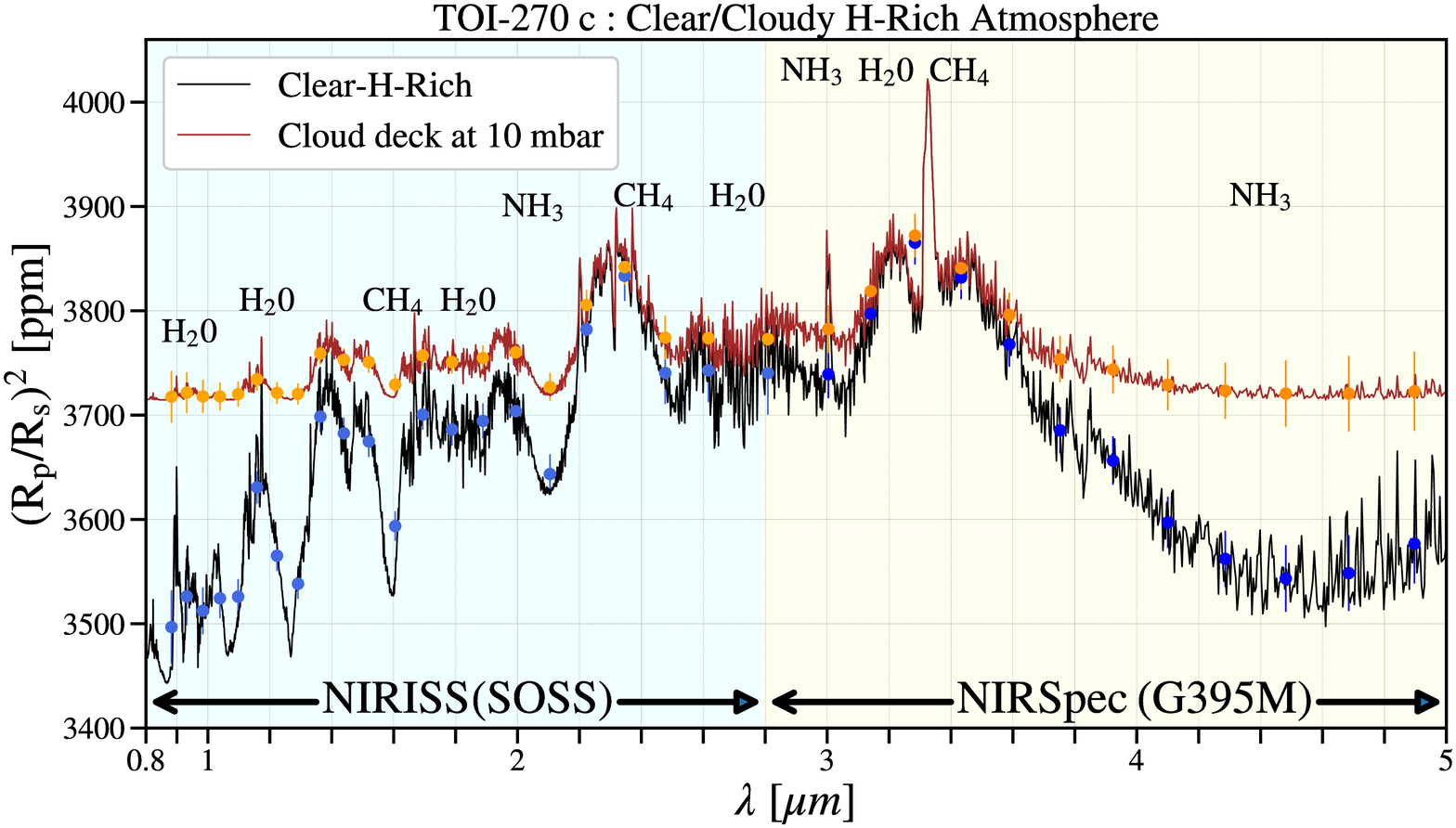}
			}%
			\subfigure{%
				
				\includegraphics[width=0.5\textwidth]{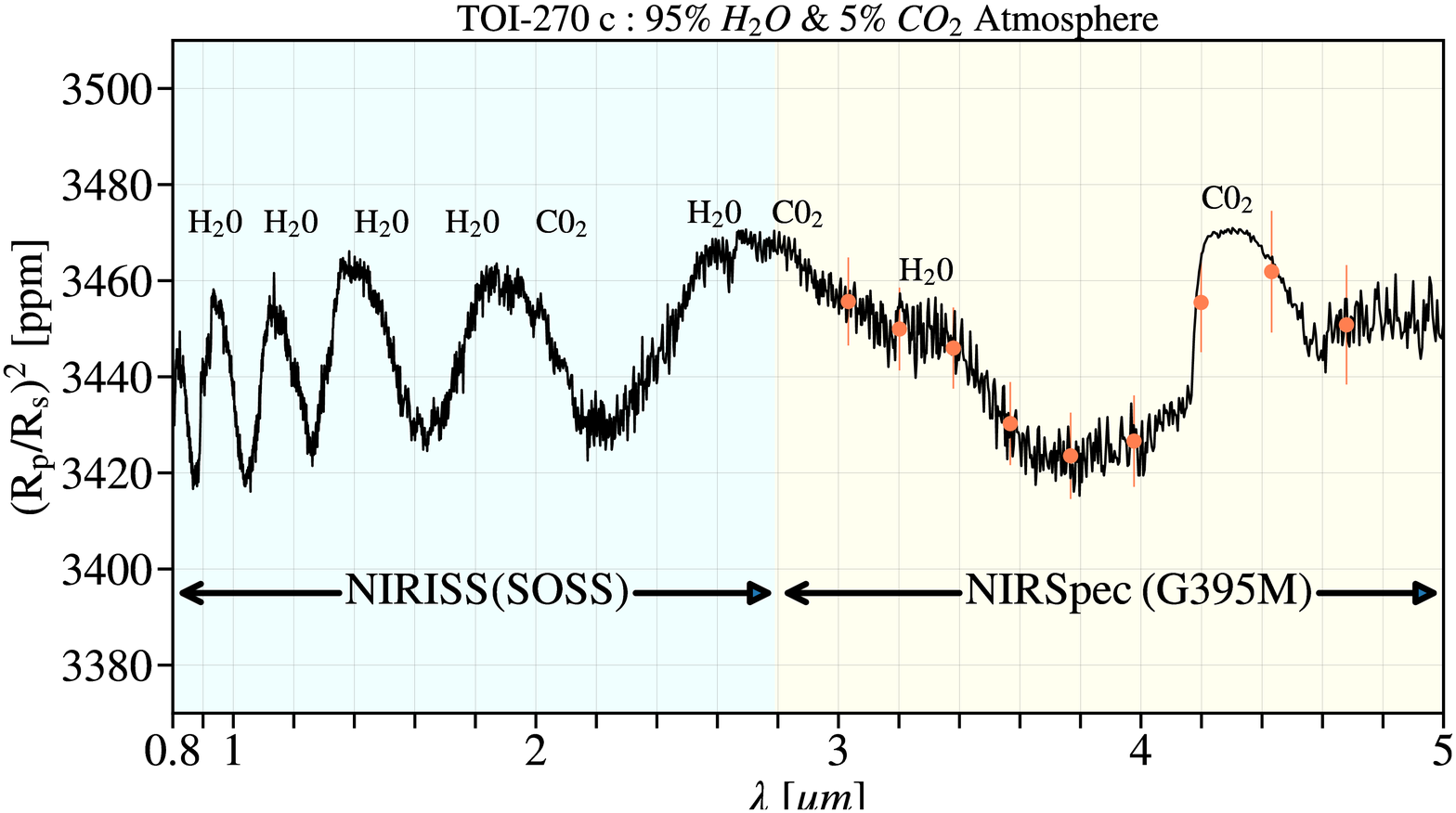}
			}\\ 
			\subfigure{%
				
				\includegraphics[width=0.5\textwidth]{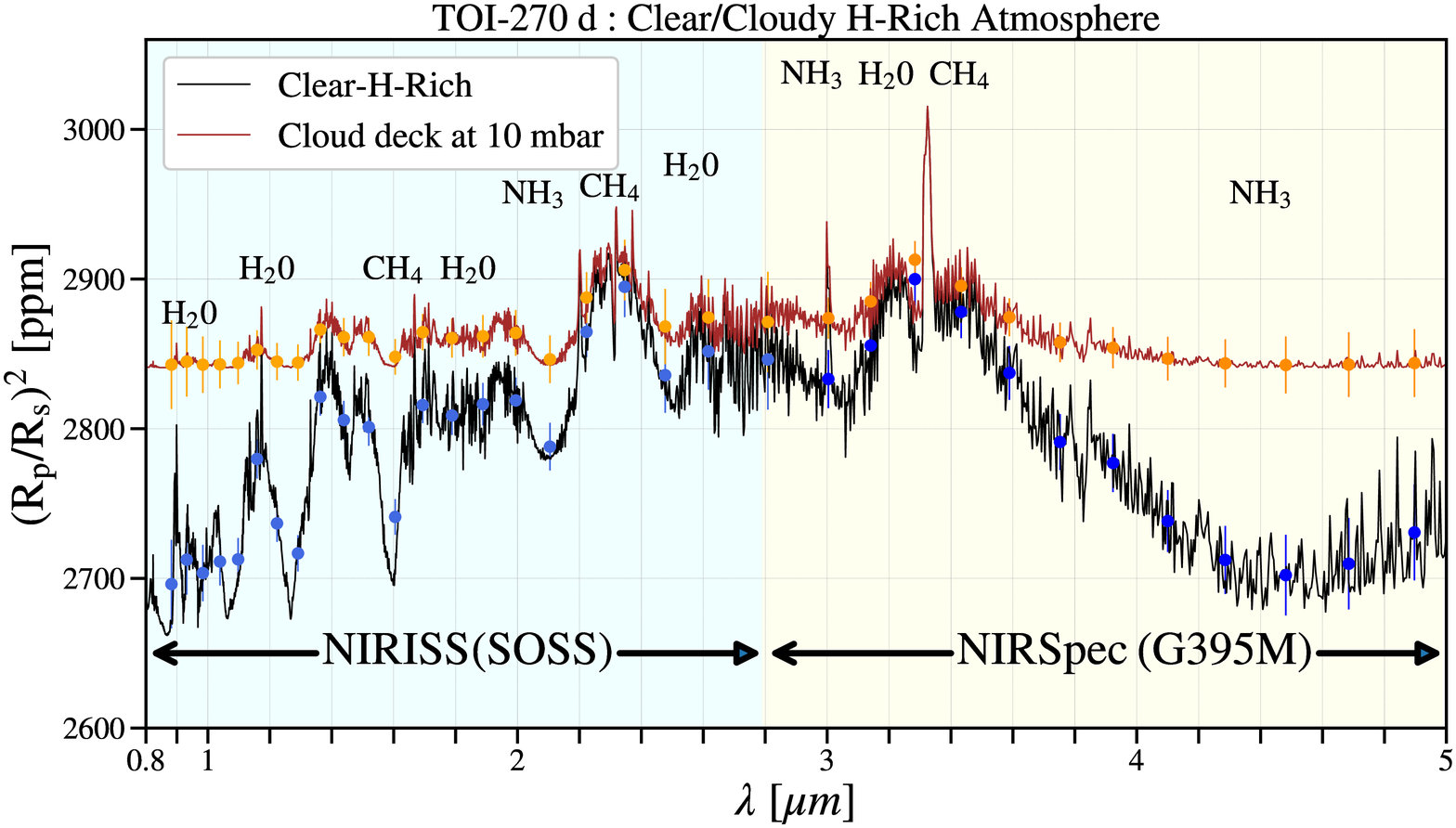}
			}%
			\subfigure{%
				
				\includegraphics[width=0.5\textwidth]{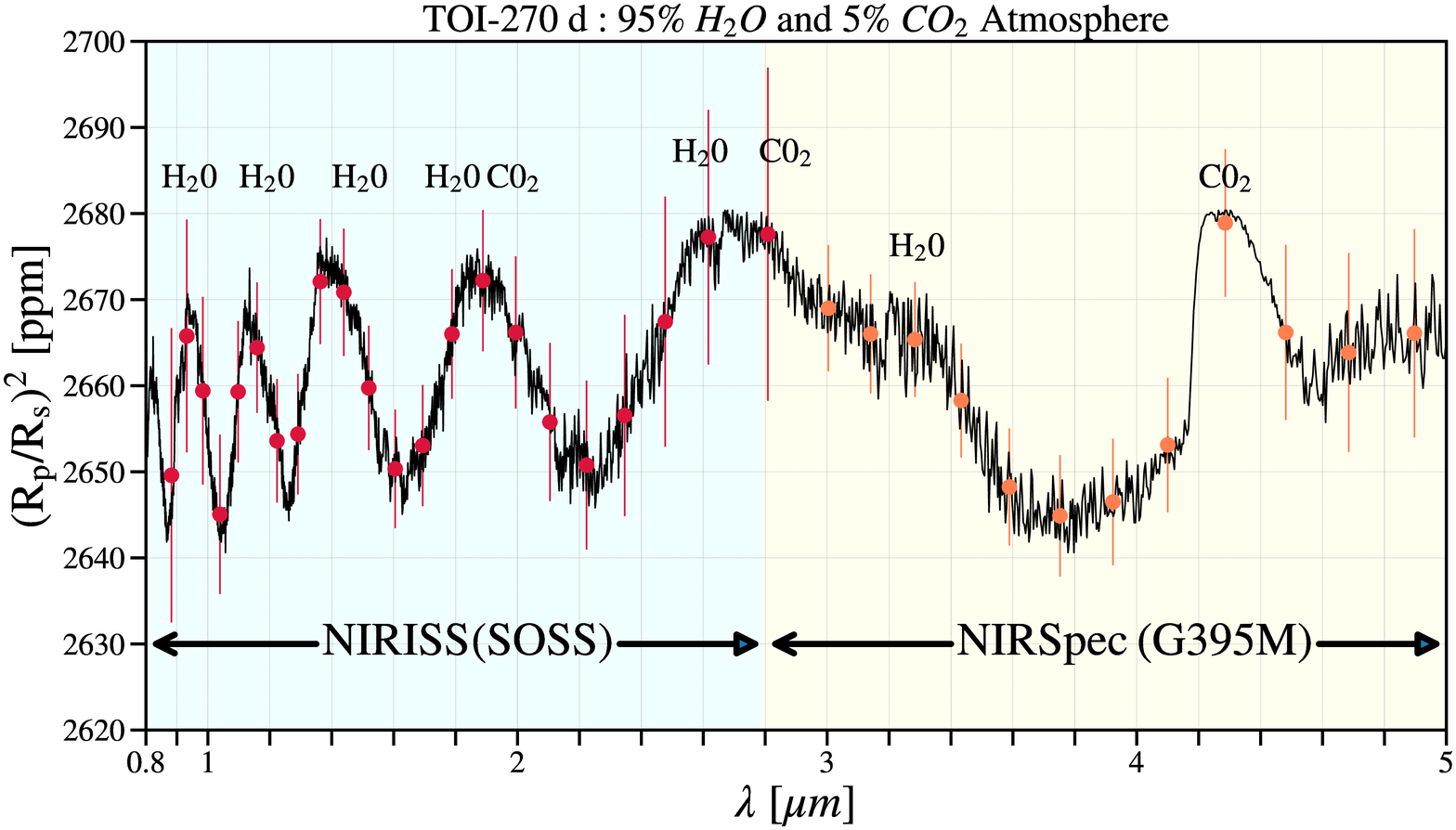}
			}%
		\end{center}
    \caption{ Simulated JWST transmission spectra of TOI-270c and d for different possible atmospheric compositions. Transmission spectrum of a H-rich atmosphere shown with error bars are calculated for 1 transit with NIRISS/SOSS which is sufficient to unambiguously detect that atmosphere with a <SNR> = 24 and <SNR > = 18 for each TOI-270c and d resp, for the clear case and with a <SNR> = 5 for the cloudy case. While the transmission spectra of the water-dominated atmosphere are calculated for 2 and 3 transits with NIRISS/SOSS to achieve a <SNR> = 5 for both planets resp. Transmission spectra calculated for 1 transit with NIRSpec/G395M is sufficient to achieve a <SNR> = 17 and <SNR > = 12 for the clear H-Rich case for both planets resp, while for the cloudy case, 1 and 2 transits are sufficient to get a <SNR > = 5 for TOI-270c and TOI-270d resp. For the water-dominated atmosphere, the transmission spectra are calculated for 5 and 7 transits with NIRSpec/G395M to achieve a <SNR> = 5 for both planets resp. The spectra are binned to a Resolution of 10.
    }%
   \label{jwst}
\end{minipage}
\end{figure*}

We have chosen to simulate the $\lambda$ = 0.8-11 $\mu$m transmission spectra of the modeled planets. This region of the spectrum shows the dominant absorption features for the major carbon-, oxygen-, and nitrogen-bearing species. Using the JWST instrument noise simulator PandExo \citep{bat}, we generate synthetic transmission spectrum observations of our models for each TOI-270c and d planet. As inputs for PandExo, we use a stellar spectrum that is taken from the PHOENIX model atmosphere library \citep{huss}, with an effective temperature of T = 3386 K, metallicity of [Fe/H] = -0.17, and surface gravity log g = 4.88, normalized to the J band magnitude of TOI-270 (Tab. \ref{para}).
We model observations of each planet and each atmospheric composition with the NIRISS/SOSS mode, the NIRSpec G395M mode and the MIRI/LRS mode. Table \ref{mode} summarizes the instrument modes considered in this study. Using these modes one obtains a close to complete spectral coverage between 0.8 and 11 $\mu$m. However, since JWST can only observe with one instrument at a time, 3 separate observations are required to obtain the full wavelength range.
In terms of instrument parameters, we set the saturation level to be 80\% of the full well capacity for all instrument modes and for the noise floor, we adopt a value of 20 ppm for the NIRISS following \citet{roch}. For the NIRSpec, we assume a noise floor value of 75 ppm as it is expected to be below 100 ppm \citep[see][]{ferr} and we set the MIRI noise floor value to 40 ppm, because the values adopted in the existing literature range from 30 to 50 \citep[see][respectively]{beich, green}. For target exposure times per transit we use the full transit durations for the TOI-270 planets from \citet{max}, and for photometry and spectroscopy noise calculations we assume that an equal amount of time is spent observing in transit versus out of transit. An example of simulated spectra for each planet is shown in Figs. \ref{jwst} \& \ref{miri}.\\
\begin{table}
	\centering             
	\caption{JWST instrument modes selected in this work}
	\label{mode}     
	\begin{tabular}{cccc} 
		\hline \hline \\
		
		Instrument & Mode & $\lambda$ ($\mu$m) & R ($ \dfrac{\Delta\lambda}{\lambda}$)  \\
		\hline \hline \\
		NIRISS & SOSS / GR700XD & 0.8-2.8 & 700  \\
		
		\hline \\
		NIRSpec & G395M & 2.9-5.0 & 1000 \\
		
		\hline \\  
		MIRI & slitless / LRS prism  & 5.0-11.0 & 100 \\  
		\hline \\    
		
	\end{tabular}
\end{table}
\begin{table}
	\centering 
	\small
	\caption{Number of TOI-270c and d transits necessary to rule out a featureless spectrum with < SNR > = 5 for the different plausible TOI-270c \& d atmospheres using different JWST instruments and modes. For the H-rich Atmosphere, only 1 transit with NIRISS (SOSS) and NIRSpec (G395M) is sufficient to get a higher <S/N> ($\ge$ 12) for both planets.}
	\label{nta}
	\begin{tabular}{ccccc}
		
		\hline \hline \\
		&	& NIRISS & NIRSpec & MIRI  \\ 
		&	& SOSS & G395M & LRS \\
		\hline \hline \\
		\textbf{TOI-270c}  & H-Rich &  1 & 1 &  1 \\ \\
		
		&   Cloudy H-Rich &  1 & 1 &  2 \\ \\
		
		&	Water-Rich 	&  2 &  5 &  25 \\ \\
		
		\hline \\
		
		\textbf{TOI-270d} & H-Rich &  1  & 1  &  2 \\ \\ 
		
		&	Cloudy H-Rich & 1  &  2  &  6 \\ \\ 
		
		& Water-Rich & 3 &  7 & 26 \\ \\ 
		\hline \\
		
	\end{tabular}
\end{table}

\section{Results}
\subsection{Detectability Assessment}
Here, we assess the JWST observations needed to detect the presence of an atmosphere for the TOI-270c \& d planets (\ref{m.d}). We then address the detectability of individual molecules within the TOI-270c \& d  planets spectra. For each atmospheric model and JWST instrument considered, we employ the procedure used in \citet{jacob} \& \citet{morley} for transit geometry. We run the PandExo JWST noise model across a grid in number of transits from 1-100, which is sufficient to establish a simple SNR scaling relationship and we determine the signal-to-noise on the difference between the model spectrum and the fiducial spectrum, which is a featureless spectrum for the case of detecting atmospheres and a spectra that are missing contributions from individual molecules for the case of detecting specific molecules. We then calculate the total expected signal-to-noise and determine the number of transit observations required to detect spectral features, such that the <SNR> is $\ge$ 5 \citep[for more details about the approach used, see][]{jacob}. This method for assessing the detectability of individual molecules neglects the complicated degeneracies that arise between molecular species that have overlapping absorption bands, and how these degeneracies may propagate uncertainty into the retrieval of individual molecules from the spectrum. However, this detectability approach provides useful indications about the observational cost (number of transits) required to make various atmospheric and molecular discoveries with JWST across a range of different atmospheric conditions and observing modes that would otherwise be computationally prohibitive to assess within an atmospheric retrieval framework. The full results of our simulations on the detectability and characterization of the TOI-270c \& d exoplanet atmospheres using JWST instruments are listed in Tab. \ref{nta} \& \ref{ntmw}.
\begin{table*}
	\centering 
	\caption{Number of TOI-270c and d transits necessary to detect molecules with transmission spectroscopy with < SNR > $\geq$ 5 for different plausible atmospheres using various JWST instruments and modes. For each molecule, the <SNR> achieved is listed in parentheses next to the number of transits required.}
	\label{ntmw}	
	\begin{tabular}{ccccc}
		\hline 
		\hline \\
		&	& NIRISS & NIRSpec & MIRI  \\ 
		&	& SOSS & G395M & LRS \\
		\hline \hline \\
		\textbf{TOI-270c} 	
		& H-Rich & $^{a}$ 1(10) ; $^{b}$1(19) ; $^{c}$1(18)  & $^{a}$2(5) ; $^{b}$1(26) ; $^{c}$1(5) &  $^{a}$2(5) ; $^{b}$1(5) ; $^{c}$2(5) \\ \\
		&   Cloudy H-Rich & $^{a}$2(5) ; $^{b}$1(7) ; $^{c}$2(5) & $^{a}$21(5) ; $^{b}$1(10) ; $^{c}$3(5) & $^{a}$6(5) ; $^{b}$2(5) ; $^{c}$4(5) \\ \\
		&	Water-Rich 	&$^{a}$2(5) ; $^{d}$51(5) & $^{a}$5(5) ; $^{d}$15(5) &  $^{a}$25(5) ; $^{d}$ > 100(5)\\ 
		\hline \\ 
		\textbf{TOI-270d} 
		& H-Rich & $^{a}$1(10) ; $^{b}$1(12) ; $^{c}$1(12) & $^{a}$3(5) ; $^{b}$1(18) ; $^{c}$2(7)  &  $^{a}$3(5) ; $^{b}$2(5) ; $^{c}$4(5)\\ \\
		&	Cloudy H-Rich & $^{a}$2(5) ; $^{b}$2(5) ; $^{c}$2(5) & $^{a}$4(5) ; $^{b}$2(5) ; $^{c}$4(5) & $^{a}$10(5) ; $^{b}$7(5) ; $^{c}$11(5)  \\ \\
		& Water-Rich 	&$^{a}$3(5) ; $^{d}$46(5) & $^{a}$7(5) ; $^{d}$19(5) &  $^{a}$26(5) ; $^{d}$ > 100(5) \\ 		
		\hline  \\
	\end{tabular} 
\newline	\footnotesize{\textbf{Notes}. $^{a}$ $H_{2}O$; $^{b}$ $CH_{4}$; $^{c}$ $NH_{3}$; $^{d}$ $CO_{2}$ \\  Number of transits for each molecules with the SNR ($N_{tran}$(<SNR>)) } 
\end{table*}
\subsection{Detectability of an Atmosphere}
\begin{figure*}
	\begin{minipage}{1\linewidth}
		\begin{center}
			\subfigure{%

				\includegraphics[width=0.5\textwidth]{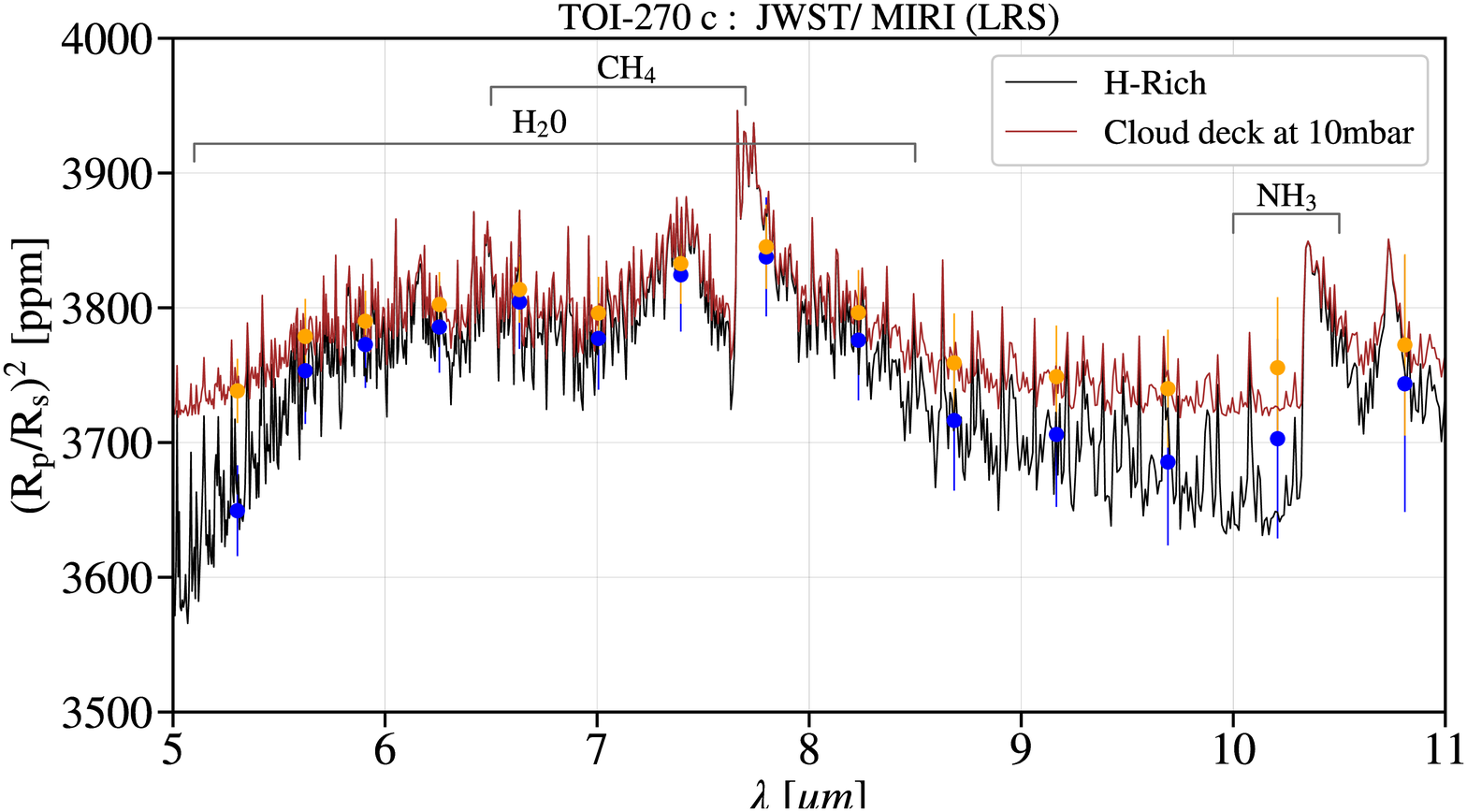}
			}%
			\subfigure{%

				\includegraphics[width=0.5\textwidth]{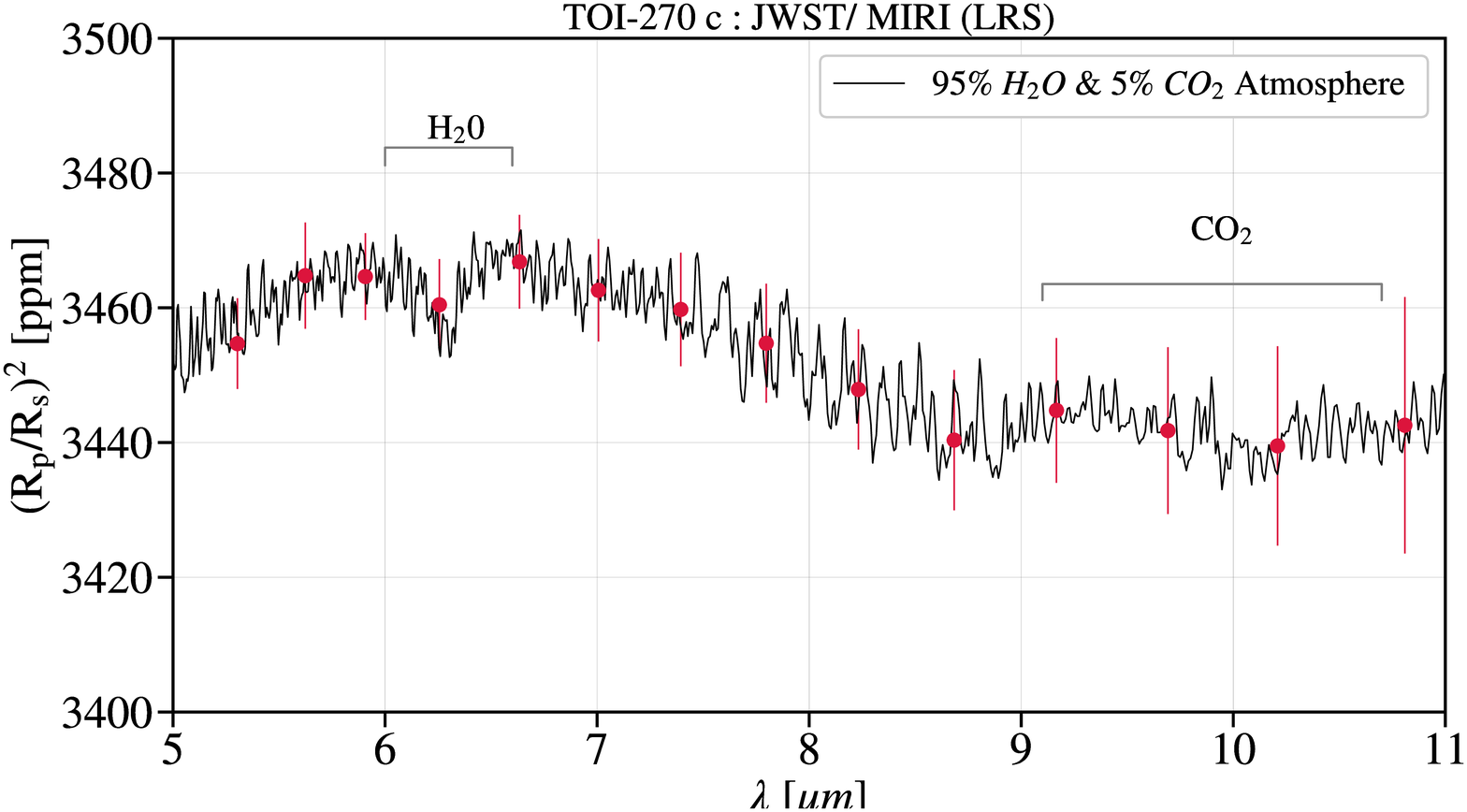}
			}\\ 
			\subfigure{%

				\includegraphics[width=0.5\textwidth]{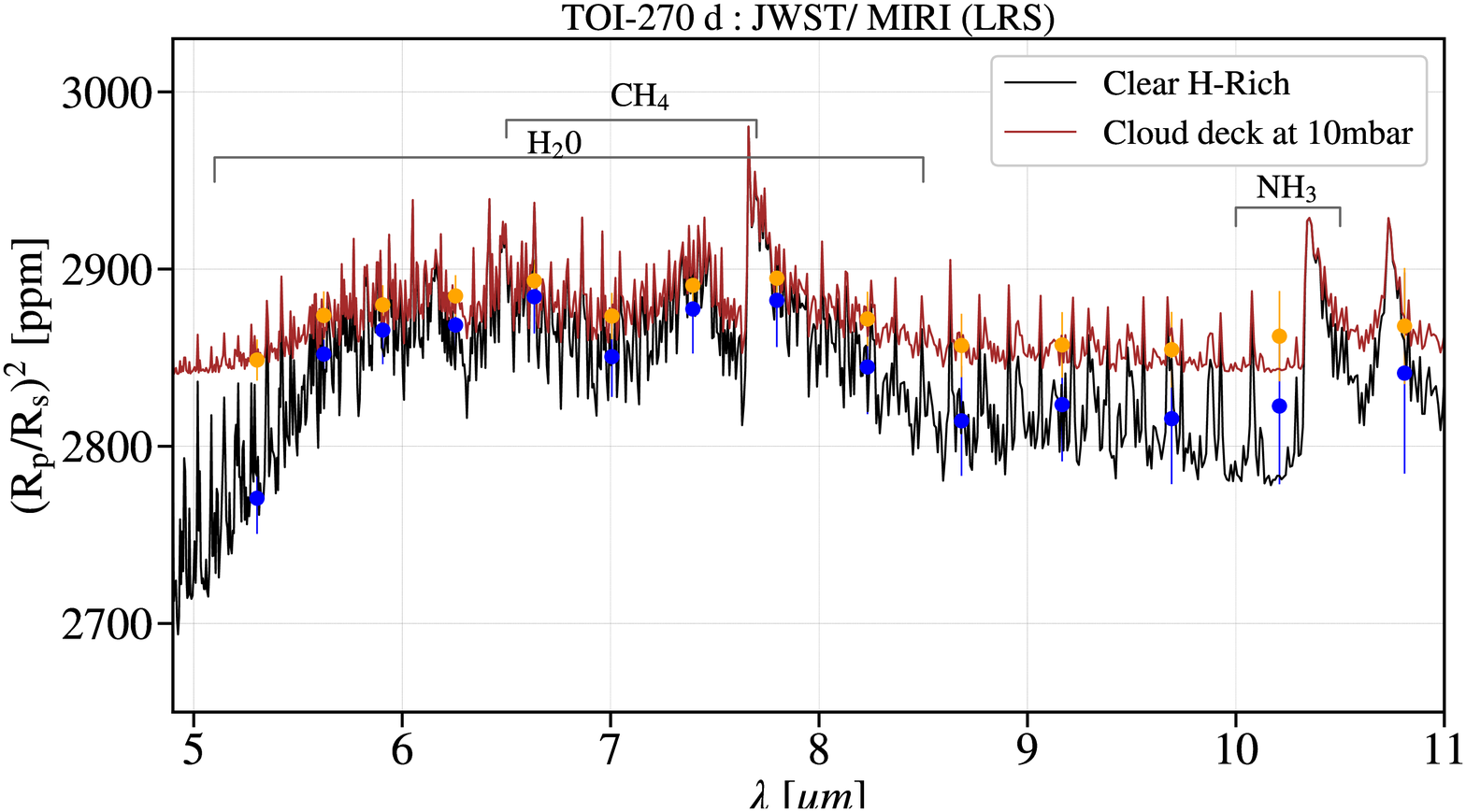}
			}%
			\subfigure{%

				\includegraphics[width=0.5\textwidth]{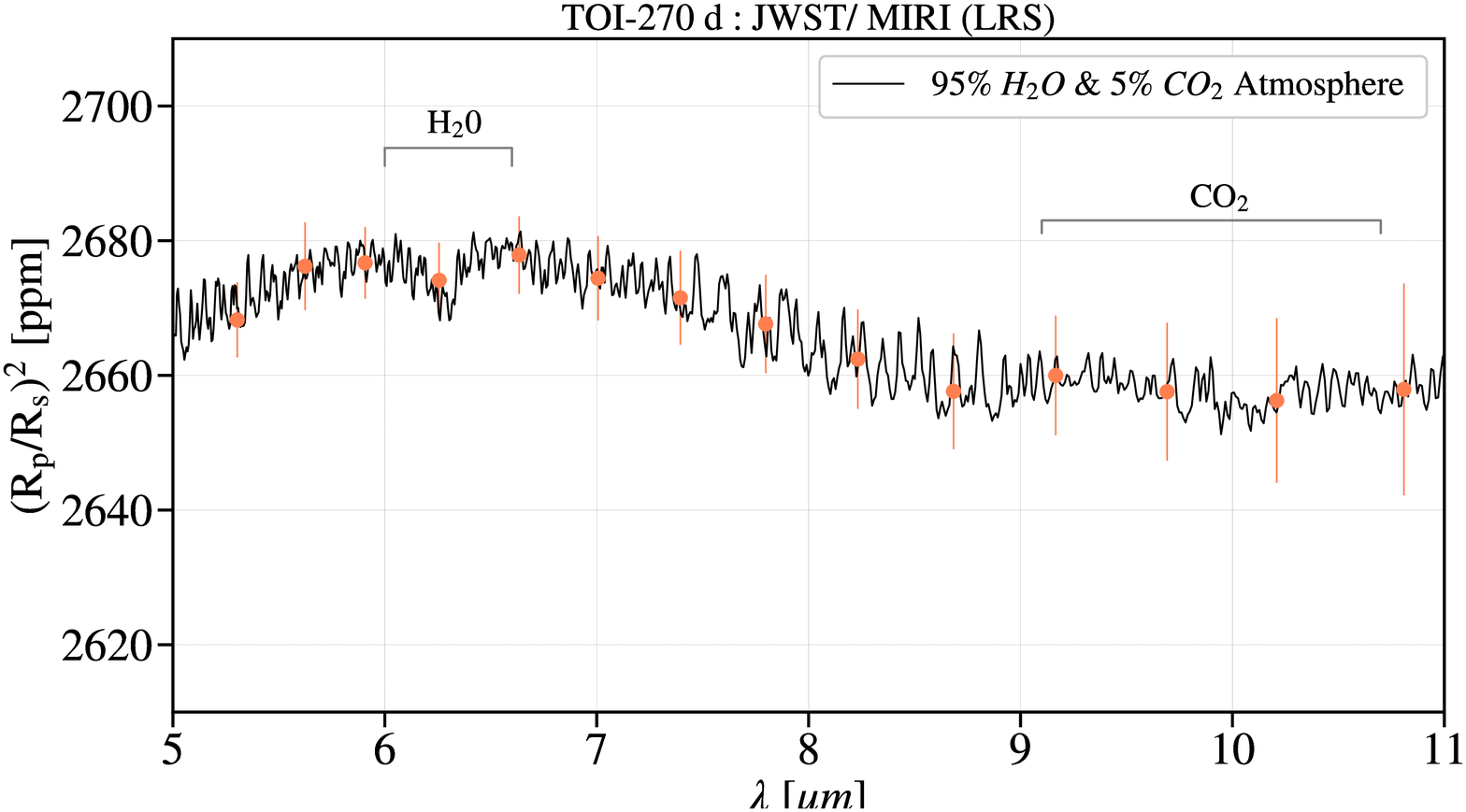}
			}%
		\end{center}
		\caption{%
			Simulated JWST transmission spectra of TOI-270c and d for different possible atmospheric compositions with JWST/MIRI (LRS) mode. The transmission spectra are shown with error bars calculated for 1, 2 and 25 transits to achieve a <S/N> of 5 for the clear and the cloudy H-rich atmospheres, and the water-dominated atmosphere, resp, for TOI-270c. For TOI-270d, the transmission spectra are calculated for 2, 6 and 26 transits.}%
		\label{miri}
	\end{minipage}
\end{figure*}

Here, we present our results on the detectability of the TOI-270c \& d planet atmospheres using transmission spectroscopy with JWST. We identify the optimal JWST mode for detecting the presence of TOI-270c \& d planet atmospheres. Figures \ref{jwst} and \ref{miri} show an example detection of absorption features in the transmission spectrum of TOI-270c \& d for different plausible scenarios and JWST modes (NIRISS/SOSS, NIRSpec/G395M and MIRI/LRS). The synthetic data are shown binned to a resolution of R = 10 for NIRISS/SOSS and NIRSpec/G395M and R= 14 for  MIRI/LRS, however the featureless spectrum was ruled out at the native resolution of the different modes (see Tab. \ref{mode}). \\
For TOI-270c, detecting the absorption features in the transmission spectrum assuming it possesses a 10 mbar, cloudy, H-rich atmosphere would require only 1 transit with the NIRISS (SOSS) \& NIRSpec (G395M) modes and 2 transits with the MIRI (LRS) mode, which we find to be sufficient to rule out a featureless spectrum with <SNR> = 5. While, we could obtain  <SNR> = 5 detection of spectral features for TOI-270d planet in 1, 2 and 6 transits with NIRISS (SOSS), NIRSpec (G395M) and MIRI (LRS) modes resp (see Tab. \ref{nta}). $CH_{4}$ and $H_{2}O$ absorption features drive the detectability of this atmosphere and are apparent in the synthetic data for both TOI-270c \& d (Fig. \ref{jwst}).\\
The water-dominated atmosphere would require more transits then the cloudy-H-rich one. This is due to its higher mean molecular weight that affect the atmospheric scale height and therefore the spectral features depth. To detect the absorption features of this atmosphere for TOI-270c, 2, 5 and 25 transits would be needed to rule out a featureless spectrum with <SNR> = 5 with NIRISS/SOSS, NIRSpec/G395M and MIRI/LRS modes, respectively, and 3, 7 and 26 for TOI-270d.\\
As can be seen in Tab. \ref{nta}, the detection of the TOI-270d atmosphere requires more transits than TOI-270c, largely due to its low temperature and therefore smaller atmospheric scale height. Moreover, the disparity in the detectability appears consistent with the different masses used for each planet. Planet mass affects the atmospheric scale height (via the surface gravity) and therefore the size and detectability of molecular features in a transmission spectrum. Fig. \ref{mr} shows how planet d is assumed to have a higher density than c, and therefore, also a higher gravity.
\begin{figure}
	\begin{minipage}{0.98\linewidth} 
		\centering
		\includegraphics[width=\columnwidth]{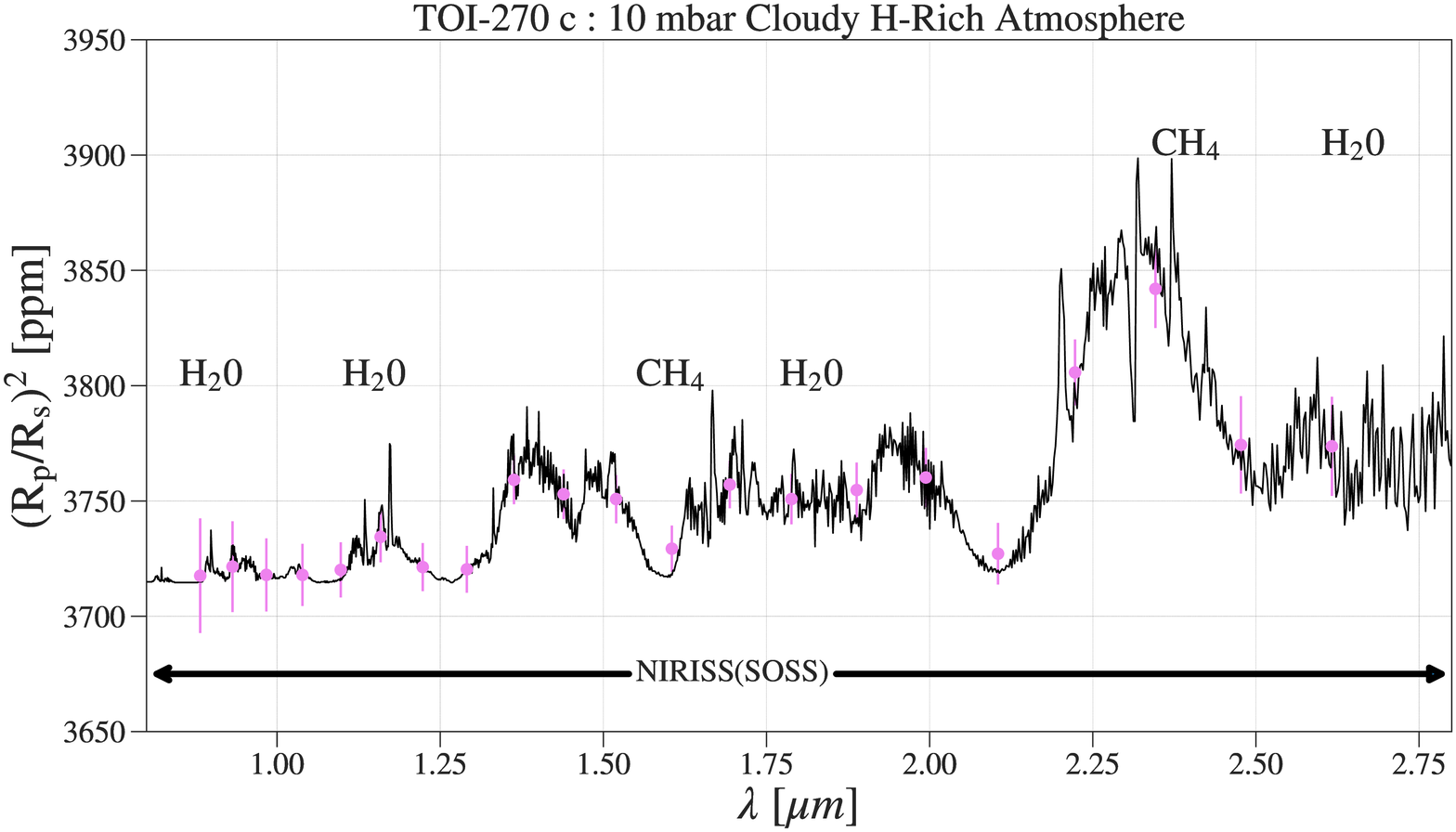}
	\end{minipage}
	\begin{minipage}{0.98\linewidth} 
		\centering
		\includegraphics[width=\columnwidth]{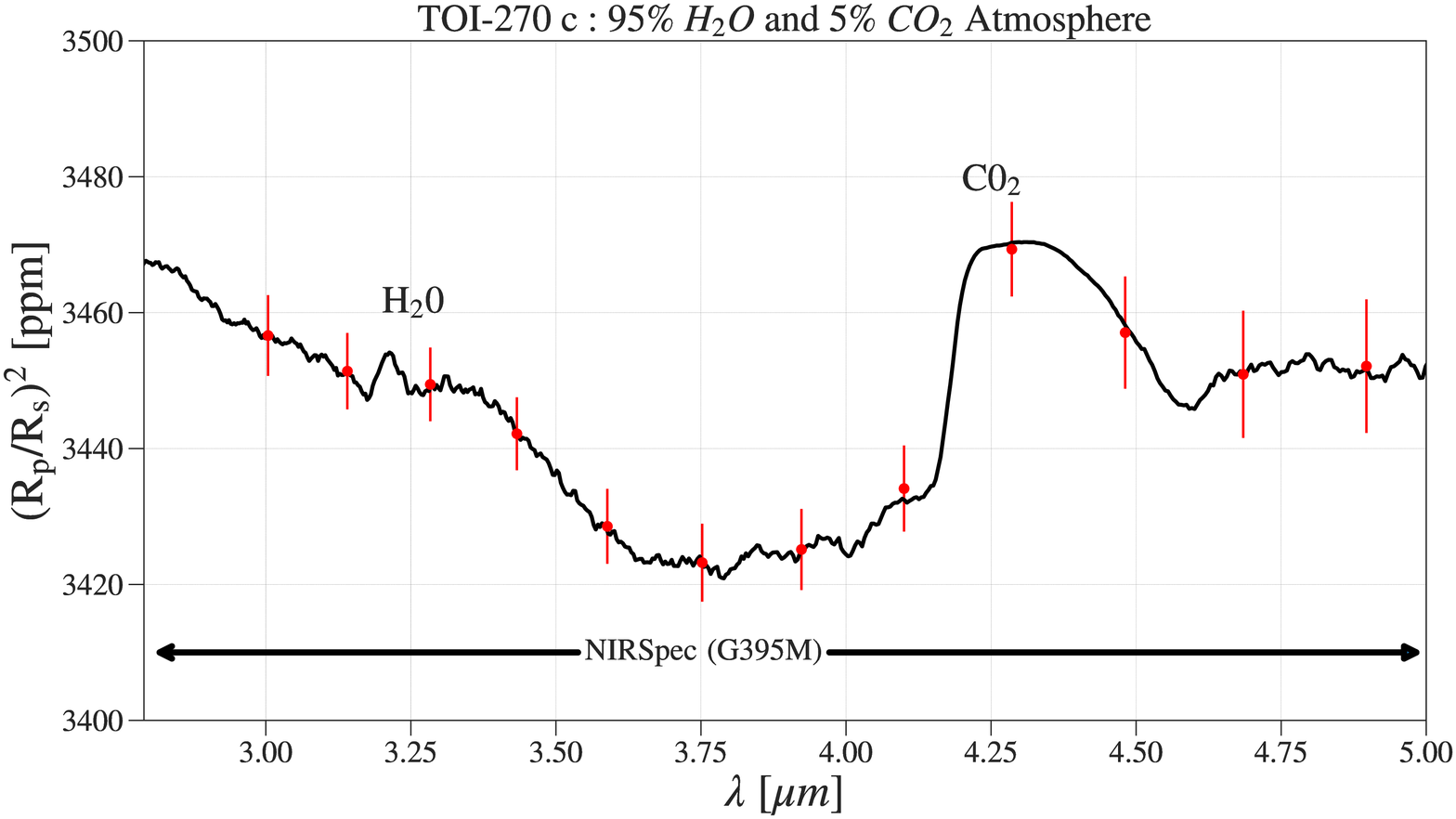}
	\end{minipage}	
	\caption{Theoretical transmission spectra of TOI-270c assuming two different atmospheric compositions with modeled noise for JWST observations. Top: Transmission spectrum of a 10 mbar cloudy H-rich atmosphere shown with error bars calculated
		for 2 transits with NIRISS (SOSS) sufficient for <SNR> = 5 on the $H_{2}O$ features. Bottom: Transmission spectrum of a water-dominated atmosphere shown with error bars calculated for 15 transits with NIRSpec (G395M) sufficient for <SNR> = 5
		on the $CO_{2}$ features.}
	\label{cl-wa}
\end{figure}
\subsection{Detectability of Individual Molecules}
We now present the sensitivity of each JWST instrument listed in Tab. \ref{mode} to each gas in the TOI-270 models transmission spectra discussed in Section \textcolor{blue}{\ref*{m.d}}. Tab. \ref{ntmw} list the molecules for which JWST could detect (<SNR> $\geq$ 5) that molecule's contribution to the spectrum, and the number of transits to do so. For each molecule, the <SNR> achieved is listed in parentheses next to the number of transits required along with a footnote identifying which molecule the observation corresponds to.\\
Fig. \ref{cl-wa} shows an example detection of absorption features in the transmission spectrum of TOI-270c for both 10 mbar cloudy H-rich atmosphere and water-dominated atmosphere with NIRISS (SOSS) and NIRSpec (G395M) respectively. The models are shown with calculated error bars that correspond to the amount of JWST observing time that would be required to detect specific molecules (H$_2$O and CO$_2$) in the given spectra. \\
The presence of $CH_{4}$ dominates the detectability of the clear and cloudy H-rich atmospheres with JWST. Even a small amount of $CH_{4}$ (e.g. $\sim$ 5ppm for the H-rich atmosphere) can saturate the strong 2.3, 3.3 and 7.8 $\mu$m $CH_{4}$ absorption features and lead to the detection of both the atmosphere and $CH_{4}$. As a result, the number of transits necessary to detect spectral features in a transmission spectrum of clear/cloudy H-rich atmosphere, are close to the number of transits necessary to detect $CH_{4}$ with the different JWST instruments listed in Tab. \ref{mode} and for both planets (see Tab. \ref{ntmw}). \\
Detecting water in the atmospheres of the TOI-270c \& d planets may help to constrain evolutionary scenarios. The presence of water may be readily detectable for TOI-270c \& d with the NIRISS (SOSS), in only 1 transit with a relatively high SNR, if they possess clear H-rich atmosphere. When we include clouds at 10 mbar in that atmosphere, the SNR on the spectral features drops which increases the required JWST time to detect water from 1 to 2 transits with NIRISS (SOSS)(Tab. \ref{ntmw} \& Fig. \ref{cl-wa}). The 2 $\mu$m $NH_{3}$ would only require one transit with NIRISS (SOSS) to be detected at SNR > 5 in the clear H-dominated atmosphere, but two transits may be required if the 
planets possess 10 mbar cloud coverage for both TOI-270c \& d.\\
In water-dominated atmosphere, $H_{2}O$ may be detectable using the NIRISS (SOSS) mode with only 2 and 3 transits for both TOI-270c \& d resp. We find that $H_{2}O$ is easier to detect than $CO_{2}$ in the water-dominated atmosphere, but this is not necessarily that surprising since the atmosphere is mostly composed of water. 
The 4.2 $\mu$m $CO_{2}$ feature would require 15 and 19 transits to be detected at <SNR> = 5 with the JWST NIRSpec (G359M) instrument (Tab. \ref{ntmw} \& Fig. \ref{cl-wa}). The spectral features in the cloudy H-rich atmosphere are easier to characterize than those in the water-dominated atmospheres. This is consistent with the mean molecular weight of the atmosphere, which lowers the scale height, and decreases the amplitude of the molecular absorption features in the transmission spectrum. \\
We find that more transits are needed to detect molecules using the MIRI (LRS) instrument, compared to other instruments, for the different atmospheres studied here. MIRI (LRS) is much less sensitive for transmission spectroscopy than the shorter wavelength JWST instruments. This is largely because there are many more stellar photons near 1 $\mu$m than out > 5 $\mu$m where MIRI LRS operates, so more photons in the NIR will give higher S/N observations. The observational difficulty with which individual molecules may be detected in a transmission spectrum varies substantially as a function of atmospheric composition. Consequently, the gases that are both relatively easy to detect and unique to a specific atmosphere make optimal testable hypotheses for that atmospheric composition.
\section{Conclusion}
We have generated forward models of transmission spectra for the newly discovered TOI-270 sub-Neptune planets. We modeled the planets with clear and cloudy hydrogen-dominated atmospheres and water-dominated atmospheres (95\% $H_{2}O$ + 5\% $CO_{2}$). We studied the effect of C/O ratio on the H-rich atmospheres for both planets. We then performed simulations of JWST observations using the instrument modes NIRISS (SOSS) over 0.8-2.8 $\mu$m, the NIRSpec (G395M) over 2.9-5.0 $\mu$m and the MIRI (LRS) over 5.0-11.0 $\mu$m. We applied the approach of \citet{jacob} to assess the detectability of these different self-consistent atmospheric compositions for both planets with the different JWST instruments. We found : 
\begin{enumerate}
	\item Using the hydrogen abundance ($X_{H}$) and the carbon to oxygen abundance ratio ($X_{C}$/$X_{O}$) as primary parameters, we explored the effect of C/O ratio on the H-rich atmospheres, and found that when $X_{H}$ > 0.7, the atmospheres have free hydrogen and chemically behave like $H_{2}$-dominated atmospheres. If $X_{H}$ $\le$ 0.5, then $H_{2}O$ can be the dominant form of oxygen and therefore the atmosphere can be water-rich for small $X_{C}$/$X_{O}$.\\
     \item Many molecular absorption features may be detectable (at <SNR> $\ge$ 10) with JWST NIRISS (SOSS) mode in only 1 transit for the clear H-dominated atmosphere for both TOI-270c \& d. The 3.3 $\mu$m $CH_{4}$ absorption feature may be strongly detected (<SNR> $\approx$ 20) in only 1 transit with NIRSpec (G395M) for both planets. In addition, multiple observations would be needed to detect spectral features using MIRI (LRS). MIRI (LRS) is much less sensitive for transmission spectroscopy than the shorter wavelength JWST instruments. 
    \item  The presence of a cloud deck at 10 mbar decreases attainable S/N on the atmosphere by about a factor of 3 for planets with hydrogen-dominated atmospheres. Multiple transits will be required to constrain molecular abundances for the TOI-270 planets if they have clouds. Detecting water in the cloudy atmosphere will require 2 transits with NIRISS (SOSS) to achieve <SNR> = 5. In addition, \citet{green} showed that cloudy atmospheres will require full 1-11 $\mu$m spectra for good constraints.\\
    \item The $H_{2}O$ spectral feature may be readily detectable (<SNR> = 5) for TOI-270c \& d with NIRISS (SOSS) in 2 and 3 transits, respectively, if they possess a water-dominated atmosphere. The $CO_{2}$ absorption feature would require longer observations $\sim$ 15-19 transits to be detected with NIRSpec (G395M) instrument. We found that the detectability of spectral features in the 10 mbar cloudy H-rich atmosphere is easier than that in the clear water-dominated atmosphere. This is consistent with the mean molecular weight of the atmosphere, which shrinks the scale height, and decreases the amplitude of the molecular absorption features in the transmission spectrum.
\end{enumerate}
The James Webb Space Telescope (JWST), with its 58 times more collecting area than Spitzer will be able to make high SNR observations of super Earths and their atmospheres. Hosting at least three super-Earths, TOI-270 lies at a distance of a mere 22 parsecs, one of the closest planetary system discovered by TESS. The planets c \& d have a high TSM \citep{kem} $\ge$ 90 which makes them excellent candidates for follow-up atmospheric characterization with JWST and other facilities. All of these exciting features mean TOI-270, like other nearby planetary systems around bright stars, will be a great asset for exploring the nature of the atmospheres of sub-Neptunian exoplanets.
\section*{Acknowledgements}
We thank the anonymous reviewer for their comments. JLY was supported by NASA's NExSS Virtual Planetary Laboratory funded by the NASA Astrobiology Program under grant 80NSSC18K0829.
\bibliographystyle{mnras}
\bibliography{jamila-mnr}
\bsp	
\label{lastpage}
\end{document}